\documentclass[reprint, superscriptaddress, amssymb, aps, pra, a4]{revtex4-2}
\usepackage{graphicx}
\usepackage{dcolumn}
\usepackage{hyperref}
\usepackage{braket}
\usepackage{bm}
\usepackage{natbib}
\usepackage{amsmath}
\usepackage{tabularx}
\usepackage{xcolor}
\usepackage{soul}

\begin{document}

\title{Anisotropic Spatial Entanglement}

\author{Satyajeet Patil} \email{satyajeet@prl.res.in}
\affiliation{Quantum Science and Technology Laboratory, Physical Research Laboratory, Ahmedabad, India 380009.}
\affiliation{Indian Institute of Technology, Gandhinagar, India 382355.}

\author{Shashi Prabhakar}
\affiliation{Quantum Science and Technology Laboratory, Physical Research Laboratory, Ahmedabad, India 380009.}

\author{Ayan Biswas}
\affiliation{Quantum Science and Technology Laboratory, Physical Research Laboratory, Ahmedabad, India 380009.}
\affiliation{Indian Institute of Technology, Gandhinagar, India 382355.}

\author{Ashok Kumar}
\affiliation{Department of Physics, Indian Institute of Space Science and Technology, Thiruvananthapuram, Kerala, India 695547.}

\author{Ravindra P. Singh} \email{rpsingh@prl.res.in}
\affiliation{Quantum Science and Technology Laboratory, Physical Research Laboratory, Ahmedabad, India 380009.}

\date{\today}

\begin{abstract}
The photon-pairs generated through spontaneous parametric down-conversion (SPDC) possess strong correlations in their transverse position and momentum degrees of freedom. For such photon-pairs, the transverse position correlation length depends on the crystal thickness and the pump wavelength, whereas the transverse momentum correlation length depends on the beam waist and spatial coherence length of the pump. By controlling these parameters, it is possible to engineer the spatial entanglement. Here, we utilize the circular asymmetry of the pump by using an elliptical-Gaussian beam to change the degree of entanglement in transverse directions, which we call anisotropic entanglement. We show the interrelation between the degree of anisotropic entanglement and the asymmetry in the beam width. In addition, we also show that for a highly asymmetric pump beam, the entanglement along the $y$ direction completely vanishes, whereas the entanglement in $x$ direction remains intact.
\end{abstract}


\maketitle

\section{Introduction}\label{sec:intro}

The spontaneous parametric down conversion (SPDC) processes obey the laws of energy and momentum conservation, and depending upon the phase matching condition, the generated two photons, \textit{signal} and \textit{idler}, are strongly correlated in continuous as well as discrete variables. Discrete variable entanglement includes polarization \cite{aspect1982experimental} and orbital angular momentum \cite{fickler2012quantum}, and have been studied and utilized in numerous applications as quantum key distribution \cite{bennett1992quantum}, superdense coding \cite{bennett1992communication}, quantum teleportation \cite{bouwmeester1997experimental}, etc. The transverse position and momentum of the SPDC photons are the continuous spatial variables that have gained attention in recent years \cite{walborn2010spatial, bohr1935can, zhang2019influence, anwar2018direct}, and found applications in the field of quantum imaging \cite{defienne2018general, defienne2019quantum, defienne2022pixel, defienne2021full, reichert2017biphoton, ndagano2022quantum}, quantum metrology \cite{taylor2016quantum} and quantum communication \cite{achatz2022certifying, walborn2006quantum, walborn2008schemes}.

The momentum correlation length of the SPDC photon pair depends upon the beam waist, and the spatial coherence of the pump beam \cite{defienne2019spatially}. Few studies have reported the effect of the degree of spatial coherence on the momentum correlation length of the SPDC photons while keeping the same beam waist \cite{defienne2019spatially, zhang2019influence}. SPDC photon pairs produced using a symmetric Gaussian pump beam with constant spatial coherence have an equal degree of entanglement along transverse directions, which we call isotropic spatial entanglement. On the other hand, an elliptic Gaussian beam has an unequal beam waist along orthogonal transverse directions. The inequality in the beam waist influences the momentum correlation lengths and gives different momentum correlation lengths in different directions; we refer to it as anisotropic spatial entanglement.

The position correlation length is written as $\Delta(x_{i}|x_{s})$, which is also known as the conditional uncertainty and tells about the uncertainty in the measurement of position of $\textit{idler}$ photon when the position of $\textit{signal}$ photon is known. Similarly, momentum correlation length is given as $\Delta(p_{x_{i}}|p_{x_{s}})$. It has been proved that when the product of position and momentum conditional uncertainties is less than 0.5 $\hbar$, the position-momentum correlations of the system exhibit EPR-like correlations \cite{reid1989demonstration, cavalcanti2009experimental}. 

There are few methods to measure the conditional uncertainty experimentally, like raster scanning of a slit in transverse directions \cite{howell2004realization, zhang2019influence}, inversion-based interferometer \cite{bhattacharjee2022measurement}, and imaging using electron-multiplying charge-coupled devices (EMCCD) \cite{edgar2012imaging, kumar2021einstein}. The slit method confines one of the photons from the SPDC photon pair using a slit and scanning for its conjugate paired photon's position. Due to the finite slit width, the resolution is limited. For higher resolution, a slit of narrow width is required. In the inversion-based interferometric method, only the paired photons provide constructive interference, and the rest of the photons overlap destructively. With the advancement in detection technology, a weak signal can be detected and registered with quantum efficiency $>90\%$ using an electron-multiplying CCD (EMCCD) camera. Here we use EMCCD based technique to find the spatial entanglement.

This work studies and compares the spatial entanglement generated using elliptical Gaussian beams (EGB) having different asymmetry. The asymmetric pump modes were prepared using the combination of two cylindrical lenses. We experimentally measured the conditional joint detection probability (JDP) of position and momentum of SPDC photon pairs and evaluated the EPR-like correlations for the spatial entanglement by varying the ellipticity of the pump beam. This article presents a brief on the theoretical background in section \ref{sec:theory} followed by the experimental details and image analysis in section \ref{sec:Experiment}. The results are presented in section \ref{sec:result} and finally we concluded in section \ref{sec:conclusion}.

\section{Theoretical Background}\label{sec:theory}
We use one dimensional Gaussian model to describe the spatial structure of the two SPDC photons field along $x$ and $y$ directions, as given below in Eqn. \ref{eq:1} \& \ref{eq:2} respectively.
\begin{eqnarray}
    \Psi\left(x_{i}, x_{s}\right)&=&N\exp \left(-\frac{\left|x_{i}+x_{s}\right|^{2}}{4 \sigma_{x+}^{2}}-\frac{\left|x_{i}-x_{s}\right|^{2}}{4 \sigma_{x-}^{2}}\right), \label{eq:1}\\
    \Psi\left(y_{i}, y_{s}\right)&=&N\exp \left(-\frac{\left|y_{i}+y_{s}\right|^{2}}{4 \sigma_{y+}^{2}}-\frac{\left|y_{i}-y_{s}\right|^{2}}{4 \sigma_{y-}^{2}}\right), \label{eq:2}
\end{eqnarray}
where, $N$ is the normalization constant, $x_{j}$, $y_{j}$ are the transverse positions of photon, $j=\{s,i\}$. The indices $s$, $i$ represent for signal and idler photons, respectively. $\sigma_{x+}$ is the standard deviation of the first Gaussian term (Eqn. \ref{eq:1}), giving momentum correlation length of photon pair, and $\sigma_{x-}$ is the standard deviation of the second Gaussian term (Eqn. \ref{eq:1}), giving position correlation length. Similarly along the $y$-axis in Eqn. \ref{eq:2}.

The JDP of two photons $J(x_{i}|x_{s}) = |\Psi(x_{i},x_{s} )|^{2}$ versus their spatial locations on the transverse plane gives Gaussian distribution, and the standard deviation of this Gaussian distribution gives position and momentum correlation length. Theoretically, the momentum correlation length along $x$ and $y$ are given as
\begin{eqnarray}
     \sigma_{x+} \propto 1/\omega_{x}, \label{eq:3}\\
    \sigma_{y+} \propto 1/\omega_{y} \label{eq:4},
\end{eqnarray}
where $\omega_{x}$ and $\omega_{y}$ are the beam waists along $x$ and $y$ directions, respectively. Position correlation length in both transverse directions are same and independent of the pump beam parameters, except wavelength. Theoretically, it is represented as,
\begin{equation}
   \sigma_{(x,y)-}=\sqrt{\frac{\alpha L \lambda_{p}}{2 \pi}},
   \label{eq:5}
\end{equation}
where $L$ is the length of the non-linear crystal, $\alpha = 0.455$ is the adjustment constant \cite{edgar2012imaging}, and $\lambda_{p}$ is the wavelength of the pump beam. In our experiment, we used $5$ mm thick type-I BBO crystal and the theoretically obtained value of the $\sigma_{x-}=\sigma_{y-}= 12.11$ $\mu$m.

We define the product of position and momentum correlation widths along $x$ and $y$ directions as
\begin{subequations}
\begin{eqnarray}
    \Delta(x_{i}|x_{s})\Delta(p_{x_{i}}|p_{x_{s}})&=&\gamma_{x}, \quad {\rm and} \label{eq:6a}\\
    \Delta(y_{i}|y_{s})\Delta(p_{y_{i}}|p_{y_{s}})&=&\gamma_{y} \label{eq:6b}.
\end{eqnarray}
\end{subequations}
For having spatial entanglement along a given axis, $\gamma$ $<$ 0.5 $\hbar$ along that axis. The number of modes $\mathcal{N}$ along that axis, respectively, is calculated as \cite{edgar2012imaging}
\begin{equation}
    \mathcal{N}_x=\left(\frac{\omega_{x}}{\sigma_{x-}}\right)^{2}, \quad {\rm and } \quad \mathcal{N}_y=\left(\frac{\omega_{y}}{\sigma_{y-}}\right)^{2}.
    \label{eq:7}
\end{equation}

\section{Experiment}\label{sec:Experiment}

\begin{figure}[h]
    \centering
    \includegraphics[width=8.5cm]{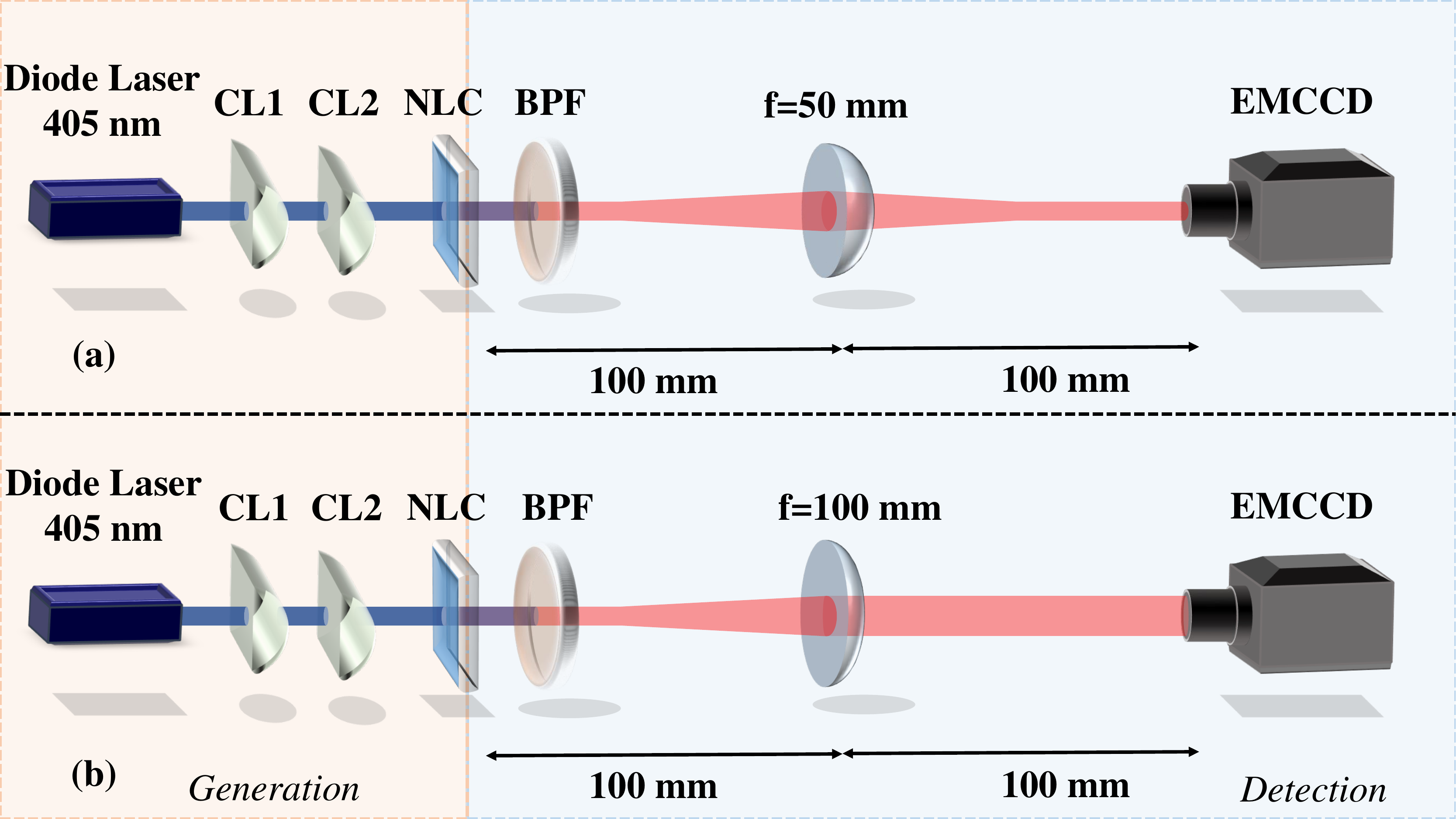}
    \caption{Experimental setup to study the position-momentum entanglement generated from an elliptical pump beam using second-order nonlinear crystal. CL1 \& CL2: cylindrical lenses; NLC: nonlinear crystal (BBO); BPF: band-pass filter. (a) Near-field configuration, (b) Far-field configuration.}
    \label{fig:setup}
\end{figure}

A schematic of the experimental setup is shown in Fig. \ref{fig:setup}. We used a 10 mW continuous-wave diode laser (TOPTICA, TopMode-405-H\_40154) of the wavelength 405 nm. The shape of the pump beam is modified to an EGB using combinations of two cylindrical lenses. A type-I non-linear BBO crystal of thickness 5 mm is kept in the pump beam path at the focal point of the second cylindrical lens in such a way that it emits collinear SPDC output at 810 nm. Throughout the experiment, the distance between crystal and EMCCD was fixed at 200 mm. To distil the SPDC output from the pump beam, we used a long pass filter (Edmund Optics, high performance long-pass Filter at 500 nm) after the BBO crystal and one more band-pass filter at 810$\pm$5 nm (Edmund Optics, high performance band-pass filter) is placed in front of the EMCCD.

The plano-convex lens of focal length 50 mm is kept between the crystal and EMCCD at 100 mm from the crystal to satisfy near-field imaging configuration, Fig. \ref{fig:setup}(a). Throughout the experiment, the distance between the non-linear crystal and EMCCD was kept fixed at 200 mm. The far-field imaging is obtained by replacing 50 mm focal length lens with the 100 mm focal length plano-convex lens, as shown in Fig. \ref{fig:setup}(b). The far-field imaging maps the transverse momentum of the photons onto position coordinates of the EMCCD.

The conditional position and momentum uncertainties are estimated using the joint probability distribution functions for photon pairs obtained in the near and far-field regime, respectively. We fit the joint probability functions with Gaussian functions, and the standard deviations of the fitted functions (say, $\sigma_{x(emccd)}$ and  $\sigma_{p(emccd)}$) give a measure of conditional uncertainties. After considering the experimental parameters that are involved, we finally obtain the actual conditional uncertainties as,
\begin{subequations}
    \begin{eqnarray}
        \Delta(x_{i}|x_{s})&=&\frac{\sigma_{x(emccd)}}{M}, \label{eq:8a}\\
        \Delta(p_{x_{i}}|p_{x_{s}})&=&\frac{k\sigma_{p(emccd)}}{f}\label{eq:8b},
    \end{eqnarray}
\end{subequations}
where $\sigma_{x(emccd)}$ is the position correlation length at the EMCCD plane measured using near-field imaging with magnification $M$ (in our experiment, $M$=1), $\sigma_{p(emccd)}$ is the momentum correlation length mapped in position coordinates of the EMCCD at the far-field, $k$ is the wave-number of the SPDC photons, and $f$ is the focal length of the lens used for far-field imaging of the crystal's exit plane on to the EMCCD.

\subsection{Pump beam shaping}\label{sec:beam_shape}
The shape of the initial circular Gaussian beam is engineered using a combination of two cylindrical lenses. The pump beam is shaped into an EGB beam such that there is a mismatch in beam width along the transverse directions. Based on the mismatch in beam width ($\omega_x, \omega_y$), we define the asymmetry factor,
\begin{equation}
    \beta=\frac{\omega_{y}}{\omega_{x}}.
    \label{eq:9}
\end{equation}
Table \ref{table:1} illustrates the combinations of two cylindrical lenses used and the corresponding $\beta$ values obtained for the EGB. Note that we orient both the cylindrical lenses so that the shape of the beam modulates only along the $y$-axis while the profile along the $x$-axis remains intact. As $\beta \rightarrow 1$, asymmetry in the beam width along transverse directions disappears.

\begin{table}[h]
    \centering
    \begin{tabular}{|p{5cm}|c|}
    \hline
    \textbf{Cylindrical lens combinations, focal lengths (f1- f2)} & \textbf{$\beta$} \\ \hline
    150 mm-50 mm & 0.193 \\ \hline
    100 mm-50 mm & 0.357 \\ \hline
     75 mm-50 mm & 0.552 \\ \hline
    None & 0.833 \\ \hline
    \end{tabular}
    \caption{Different combinations of two cylindrical lenses with focal lengths f1 and f2, and the corresponding asymmetry factor $\beta$ for the pump beam.}
    \label{table:1}
\end{table}

\subsection{EMCCD settings and data acquisition}\label{sec:emccd}
To record the data, we used Andor's iXon 633 EMCCD having a pixel size of 16 $\mu$m, operating in kinetic mode at -60$^{\circ}$C. In the acquisition setup, we keep the following settings of the EMCCD identical for both position and momentum measurements: gain $\times$1000, horizontal readout rate 1 MHz with pre-amplifier gain $\times$1, vertical shift speed 3.3 $\mu$s, and vertical clock voltage amplitude as normal. We chose 64$\times$64 sensor area, encompassing the complete transverse profile of the near and far-field imaging of the SPDC output. Different exposure times, 3.5 ms for position and 5 ms for momentum, were chosen to maintain a mean number of photons per pixel ($\approx$0.004 photons/pixel). 

First, we collected the data for dark noise of the EMCCD for different exposure times, 3.5 ms and 5 ms. To measure average photons per pixel due to dark noise, we acquired 0.25 million frames with the internal shutter closed such that no light enters the EMCCD. Later, with internal shutter open, we collected the 0.25 million frames of the SPDC photons at 3.5 ms and 5 ms exposure times for the near and far-field imaging, respectively. From the BBO crystal to the EMCCD, the entire setup was covered with black anodized aluminium sheets to block noise due to stray photons. For pump beams with different $\beta$ values, we followed the same technique as discussed above. The EMCCD takes $\sim$3 hours to capture 0.25 million frames.

\subsection{Image analysis}\label{sec:image_analysis}
References \cite{edgar2012imaging, ndagano2020imaging, defienne2018general, defienne2019quantum} report the method of finding spatial correlations using EMCCD in detail. Here we explain the method that we use to find the position and momentum correlations. First, we calculate the average number of photons ($\overline{I}_{d}$) and standard deviation of the photons $\delta_{d}$ per pixel from the frames taken at 3.5 ms to quantify the dark noise of EMCCD. To get rid of the internal noise of the EMCCD, $\overline{I}_{d}$ is subtracted from all the pixels of every frame of the SPDC signal. For a pixel, if the subtraction falls below the ($\delta_{d}$) then that pixel value is assigned to zero, otherwise we keep its original value. This process is known as thresholding, and it improves the signal-to-noise ratio of the results. These thresholded frames are then used for further analysis.

Apart from the dark noise, there are other sources of noise such as stray photons, pump photons leaked from the long-pass and band-pass filters, SPDC photons with a lost partner, etc. To get rid of such noise, we utilize pixel-to-pixel intensity correlation. To find the position correlation length along the $x$ direction, the first thresholded frame is summed column-wise along the $y$ direction. This summation converts the 64$\times$64 square matrix into a 1$\times$64 row matrix. This row matrix is then multiplied by its transpose, which is 64$\times$1 column matrix. The multiplication of column and row matrix gives a 64$\times$64 matrix in which the diagonal elements correspond to the intensity correlation of pixels with themselves. The first sub-diagonal element corresponds to the intensity correlation between the photons for which the spatial separation is 1 pixel. Similarly, the second sub-diagonal element shows the intensity correlation between the pixels for which the spatial separation is two pixels and so on. The distribution of intensity correlation with the pixel separation gives the JDP. This JDP of photon-pairs can have contributions from the stray noise as discussed above. To subtract out such unwanted contributions, we follow the technique described in Ref.\cite{defienne2018general}. 

As described in Ref \cite{defienne2018general}, JDP is a function of the position of photon-pairs and is given as,
\begin{equation}
    JDP(x)= \frac{1}{N_{f}} \sum_{l=1}^{N_{f}} I_{l}\left(x_{i}\right) I_{l}\left(x_{s}\right)-\frac{1}{N^{2}_{f}} \sum_{l \neq l^{'}}^{N_{f}} I_{l}\left(x_{i}\right) I_{l^{'}}\left(x_{s}\right), 
    \label{eq:10}
\end{equation}
where $N_{f}$ is the number of frames. $I_{l}$ is the number of photons in the $l^{th}$ thresholded frame, $x_{i}$ and $x_{s}$ are the position coordinates of $idler$ and $signal$ photons, respectively.

To calculate the JDP, the first and second frames are summed along the $y$ axis to get a row matrix. Then, the row matrix of the second frame is transposed to a column matrix and multiplied by the first frame's row matrix. This creates the 64$\times$64 matrix in which the pixel-to-pixel intensity correlation between two frames is present (first term of the Eqn. \ref{eq:10}). Since both the photons from the SPDC photon pairs are simultaneously generated within the inverse bandwidth of the process, they hit the EMCCD together in a given exposure time, and there is zero probability of hitting on the two consecutive frames separately, which are acquired within a time interval larger than the inverse bandwidth of the process. Therefore, the JDP between two consecutive frames gives the JDP due to the noise and not due to the SPDC photon pairs (second term of the Eqn. \ref{eq:10}). When JDP obtained from successive frames is subtracted from the JDP obtained from the same frames, it gives the resultant JDP only due to the SPDC photon pairs. This procedure is repeated over 0.25 million frames, and the resultant JDP is averaged out, which looks like Fig. \ref{fig:jdp}(a1-c1). 

In order to obtain a quantitative estimation of position and momentum correlations, the JDP for the spatial separation between photon pairs is fitted to a sum of two Gaussian distributions. The first Gaussian distribution corresponds to the noise curve, whereas the other corresponds to the signal curve. For each Gaussian distribution, the standard deviation is calculated, and the resultant standard deviation is obtained using the following formula, 
\begin{equation}
    \mathcal{S}=\frac{\mathcal{S}_{s}\mathcal{S}_{n}}{\sqrt{\mathcal{S}_{s}^{2}+\mathcal{S}_{n}^{2}}},
    \label{eq:11} 
\end{equation}
where $\mathcal{S}_{s}$ and $\mathcal{S}_{n}$ are the standard deviations of the Gaussian distributions corresponding to signal and noise, respectively. A multiplication by the pixel size (16 $\mu$m) to the resultant standard deviation gives the position correlation between SPDC photon pairs at the EMCCD plane. As explained in  Sec. \ref{sec:Experiment}, the position correlation length at the crystal is calculated by dividing the position correlation length at the EMCCD plane with magnification factor $M$ . The JDP for momentum correlations looks like Fig. \ref{fig:jdp}(a2-c2). To calculate the momentum correlation length, the width of JDP distribution is multiplied by $f/k$, as discussed in section \ref{sec:Experiment}.

\section{Results and discussion}\label{sec:result}
\begin{figure}[h]
    \centering
    \includegraphics[width=8.8cm]{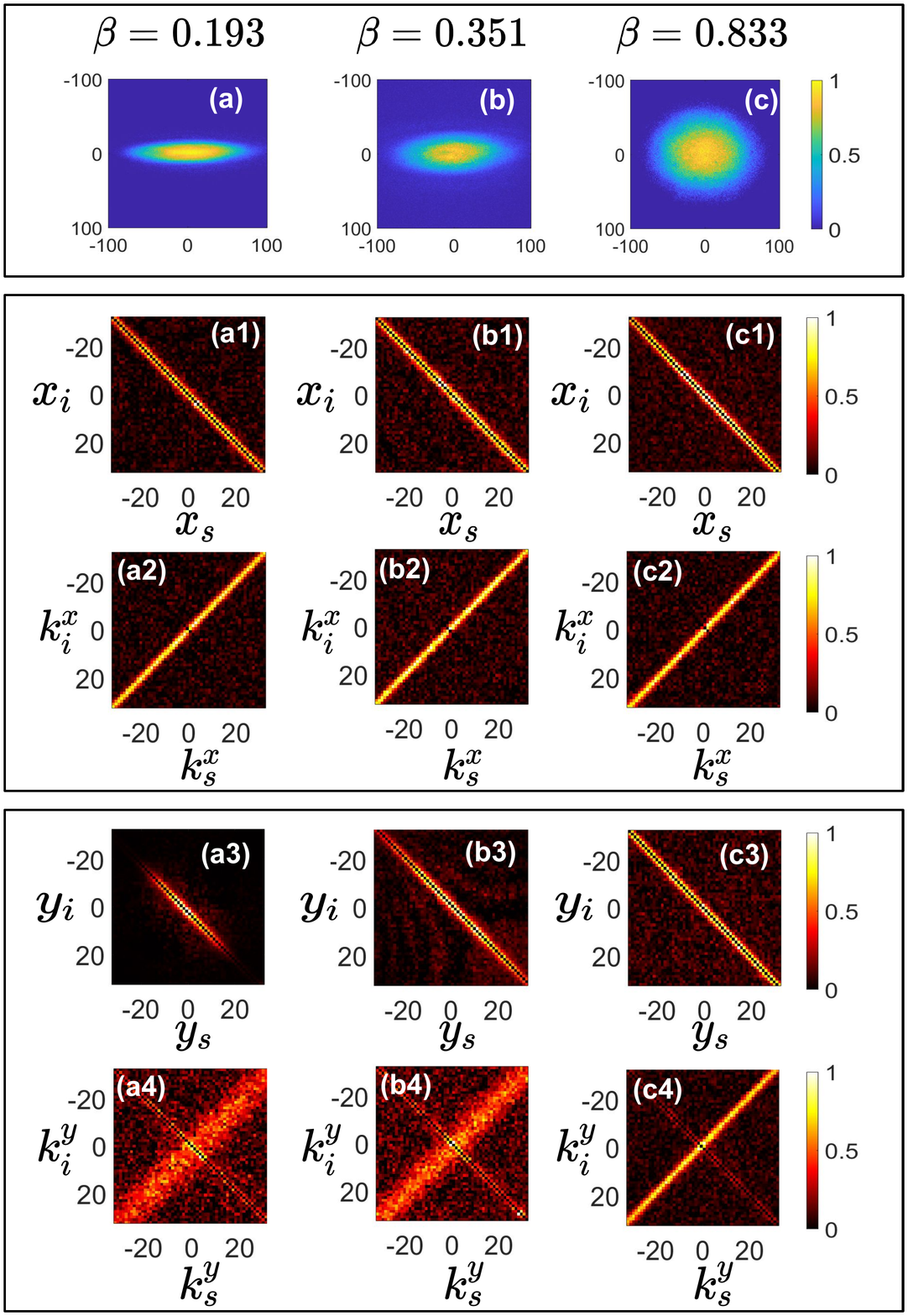}
    \caption{Intensity distribution of the pump beam and the obtained JDP for the position and momentum in $x$ and $y$ directions. (a), (b), and (c) are for different pump beams with $\beta$= 0.193, 0.351, 0.833, respectively. For $\beta$ = 0.193, (a1, a3) represent the JDP of the position along $x$ and $y$, (a2, a4) the momentum along $x$ and $y$; The same follows for $\beta$ = 0.351 and $\beta$= 0.833 in (b1-b4) and (c1-c4), respectively.}
    \label{fig:jdp}
\end{figure}

Obtained results from the experiment are shown in Fig. \ref{fig:jdp}. The first row of Fig. \ref{fig:jdp}(a-c) shows the transverse profile of the pump beams having different $\beta$ values. The color bar shows the normalized intensity distribution.

The position correlation length remains the same for different $\beta$ as it is independent of pump beam parameter, Eqn.\ref{eq:5}, and this is observed experimentally too. The width of the JDP for position remains same for the $x$ and $y$ directions as shown in Fig. \ref{fig:jdp}(a1-c1) \& (a3-c3).

For momentum correlation length along $x$ direction, width of the momentum JDP for different $\beta$ values is nearly equal because the beam width along $x$ direction is almost equal, Fig. \ref{fig:jdp}(a2-c2). Contrary to the $x$ direction, width of the momentum JDP along the $y$ direction changes with the $\beta$ value, Fig. \ref{fig:jdp}(a4-c4). For the lower values of $\beta$, the JDP width is greater than that for the higher values of $\beta$. 

\begin{figure}[h]
    \centering
    \includegraphics[width=8cm]{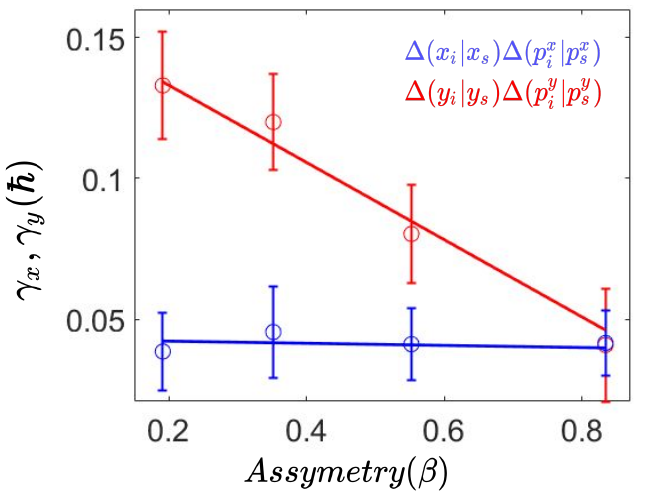} 
    \caption{The variation of $\gamma$ along $x$-axis (blue) and $y$-axis (red) against the asymmetry factor $\beta$. $\gamma_{x}$ and $\gamma_{y}$ are defined in Sec. \ref{sec:theory} has the dimensions of $\hbar$.}
    \label{fig:ent}
\end{figure}

We also calculate the variation of $\gamma$ as a function of the asymmetry, Fig. \ref{fig:ent}. It can be seen from the plot that as the asymmetry parameter goes away from one, the value of $\gamma$ increases. For different asymmetry parameters, we have shown the conditional uncertainty products and the number of modes in $x$ and $y$ directions in Table \ref{table:2}.

\begin{table}[h]
    \centering
    \begin{tabular}{|c||c|c||c|c|}
    \hline
    $\bm{\beta}$ & $\bm{\gamma_{x} (\hbar)}$ & $\bm{\gamma_{y} (\hbar)}$ & $\bm{\mathcal{N}_x}$ & $\bm{\mathcal{N}_y}$ \\ \hline
    0.193 & 0.038 & 0.133 & $\sim$ 3900 & $\sim$ 150 \\ \hline
    0.351 & 0.045 & 0.120 & $\sim$ 3500 & $\sim$ 400 \\ \hline
    0.552 & 0.080 & 0.041 & $\sim$ 3600 & $\sim$ 1100 \\ \hline
    0.833 & 0.041 & 0.040 & $\sim$ 4000 & $\sim$ 3900 \\ \hline
    \end{tabular}
    \caption{$\gamma_{x}$, $\gamma_{y}$, $\mathcal{N}_x$ and $\mathcal{N}_y$ along $x$- and $y$-direction for the four different values of $\beta$.}
    \label{table:2}
\end{table}

\begin{figure}[h]
    \centering
    \includegraphics[width=8.5cm]{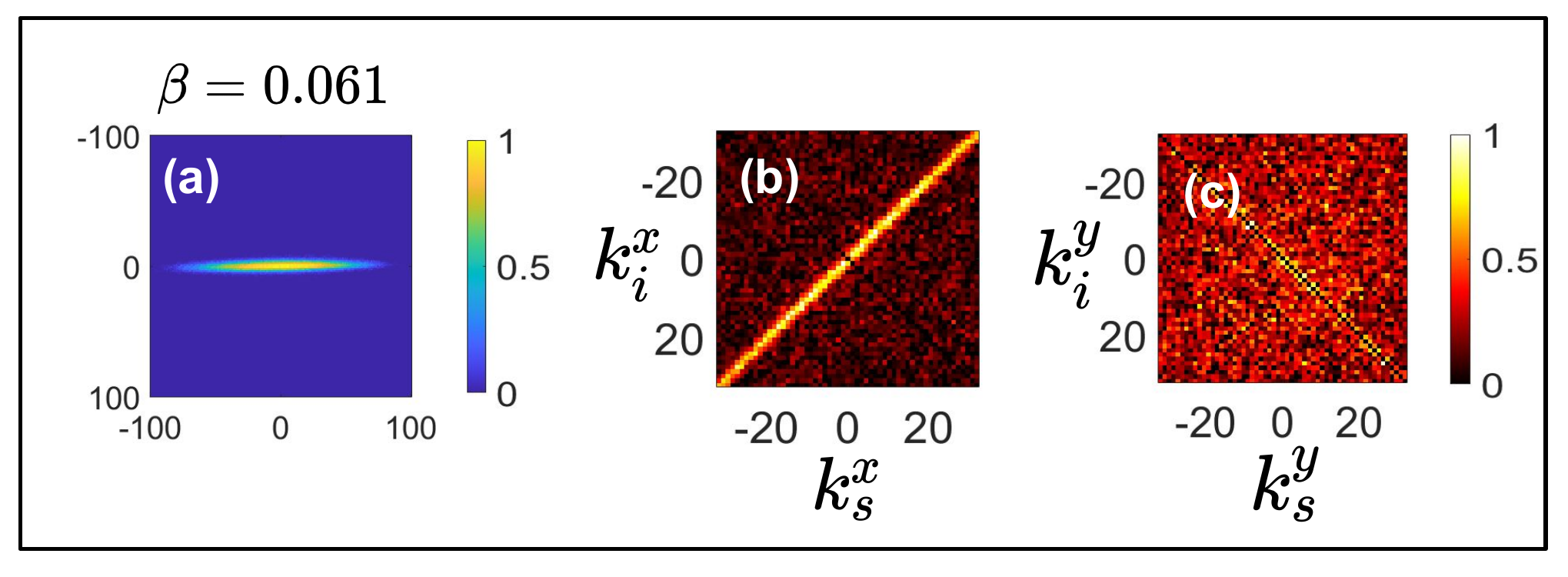}
    \caption{ (a) Intensity distribution of the pump beam with $\beta$=0.061, (b-c) are the momentum JDP for $x$ and $y$ directions, respectively. Color bar represents the normalized intensity (a) and the normalized coincidence counts (b,c).}
    \label{fig:conv}
\end{figure}

Furthermore, in the context of spatial entanglement distribution in the transverse plane, it would be interesting to know the asymmetry parameter for which the spatial entanglement is completely lost in one direction, while still existing in its orthogonal direction. In Fig. \ref{fig:conv}, we show the entanglement death in $y$ direction for $\beta$ = 0.061 while keeping it alive along x-direction. To generate the pump beam with such a low asymmetry value, we used a cylindrical lens of focal length of 200 mm to focus the EGB on the non-linear crystal. As shown in the Fig. \ref{fig:conv}(b), the width of momentum correlation along $x$ direction remains same as that in Fig. \ref{fig:jdp}(b-c), but the width of momentum correlation along $y$ direction is completely spread over entire JDP matrix, Fig.\ref{fig:conv}(c). Experimentally obtained values of $\gamma_{x}$, $\gamma_{y}$, $\mathcal N_{x}$, and $\mathcal N_{y}$ for the $\beta$=0.061 are given in the Table \ref{table:3}.

\begin{table}[h]
    \centering
    \begin{tabular}{|c||c|c||c|c|}
    \hline
    $\bm{\beta}$ & $\bm{\gamma_{x} (\hbar)}$ & $\bm{\gamma_{y} (\hbar)}$ & $\bm{\mathcal{N}_x}$ & $\bm{\mathcal{N}_y}$ \\ \hline
    0.061 & 0.037 & 0.713 & $\sim$ 3700 & $\sim$ 15 \\ \hline
    \end{tabular}
    \caption{$\gamma_{x}$, $\gamma_{y}$, $\mathcal{N}_x$ and $\mathcal{N}_y$ along $x$- and $y$-direction to observe entanglement death along one axis and high degree of entanglement in another axis.}
    \label{table:3}
\end{table}

\section{Conclusion}\label{sec:conclusion}
In conclusion, we used the combination of two cylindrical lenses to shape the elliptical-Gaussian beam with distinct beam widths along transverse directions. This asymmetry in the beam widths reflects in the momentum correlation length of the SPDC photon pairs. Since the position correlation length is independent of the pump beam parameters, the spatial entanglement is mainly affected by the momentum correlation lengths. We experimentally show that the spatial entanglement along the $y$ direction is controlled using circular asymmetry of the pump beam while the spatial entanglement remains intact along the $x$ direction. As the asymmetry in the pump beam increases, the quality of entanglement and the number of modes in $y$-direction decreases while the number of modes in $x$-direction approximately remains constant. 

We also show that for a highly asymmetric pump beam, entanglement in the $y$-direction completely vanishes, whereas the entanglement in the $x$-direction is preserved. Such anisotropic entangled states may find application in quantum communication and quantum imaging, where correlation with different strengths is required.

\section*{Acknowledgement}
We gratefully acknowledge Prof. Daniele Faccio for explaining the method of analyzing the images to obtain correlation lengths. S. Patil acknowledges Abhinandan Bhattacharjee for helpful discussion, and also acknowledges VIKRAM-100 cluster at Physical Research Laboratory, Ahmedabad. AK would like to acknowledge the support from Science and Engineering Research Board, India, through DST-SERB under Grant SRG/2019/001631.

\section*{Disclosures}
The authors declare no conflicts of interest related to this article.

\bibliography{ref}

\end{document}